\documentclass[sigconf,screen]{acmart}
\acmBooktitle{Companion Proceedings of the 34th ACM Symposium on the Foundations of Software Engineering (FSE '26), June 5--9, 2026, Montreal, Canada}
\AtBeginDocument{%
  }

\usepackage{booktabs,tabularx,threeparttable}
\definecolor{beige}{HTML}{F5F5DC}

\usepackage[most]{tcolorbox}
\usepackage{algorithm}
\usepackage{algpseudocode}
\usepackage{graphicx}
\usepackage{subcaption}
\setcopyright{acmlicensed}
\copyrightyear{2026}
\acmYear{2026}
\setcopyright{cc}
\setcctype{by}
\acmConference[PROMISE '26]{22nd International Conference on Predictive Models and Data Analytics in Software Engineering}{July 05, 2026}{Montreal, QC, Canada}
\acmBooktitle{22nd International Conference on Predictive Models and Data Analytics in Software Engineering (PROMISE '26), July 05, 2026, Montreal, QC, Canada}
\acmDOI{10.1145/3803846.3807469}
\acmISBN{979-8-4007-2584-5/2026/07}

\begin{document}

\title{Strategies for Guiding LLMs to Use Software Design Patterns: A Case of Singleton}


\author{Viktor Kjellberg}
\orcid{0009-0006-4330-2385}
\affiliation{%
  \institution{University of Gothenburg and Chalmers University of Technology}
  \city{Gothenburg}
  \country{Sweden}
}
\email{viktor.kjellberg@gu.se}
\author{Farnaz Fotrousi}

\orcid{0000-0001-5385-0381}
\affiliation{%
  \institution{University of Gothenburg and Chalmers University of Technology}
  \city{Gothenburg}
  \country{Sweden}
}
\email{farnaz.fotrousi@gu.se}
\author{Miroslaw Staron}

\orcid{0000-0002-9052-0864}
\affiliation{%
  \institution{University of Gothenburg and Chalmers University of Technology}
  \city{Gothenburg}
  \country{Sweden}
}
\email{miroslaw.staron@gu.se}

\begin{abstract}
Large Language Models (LLMs) can generate functional source code from natural-language prompts, but often fail to consistently follow higher-level architectural structures or design patterns. Since LLMs are increasingly used in software engineering, their ability to apply established design principles to generated code is crucial to the long-term success of software products. Therefore, the goal of this paper is to identify strategies for guiding LLMs to incorporate design patterns into the generated source code. We designed a computational experiment to evaluate the ability of 13 LLMs to generate code that follows the Singleton design pattern, using four prompting strategies: instructions, binary automated feedback, extensive automated feedback, and extensive feedback with few-shot prompts, in 164 Java coding challenges from HumanEval-X. 
Our results shows that the optimal strategy to guide LLMs to include design patterns depends heavily on the type of model. Still, overall, iterative binary feedback provides the best alignment with Singleton while preserving or improving the code's functionality. With guiding with instructions, Llama 3.3 generated Singleton classes in 100\% of cases and improved code functionality, increasing the number of tests passed by 34.1 percentage points. It achieved a similar result with guidance through instructions and binary feedback. Qwen 3 (8B) increased the alignment with Singleton to 99.2\% and the functionality to 58.6\% using binary feedback. Our result suggests that even simple strategies can be used to guide LLMs to use design patterns.

\end{abstract}

\ccsdesc[300]{Software and its engineering~Design patterns}
\ccsdesc[300]{Computing methodologies~Natural language processing}
\ccsdesc[100]{Software and its engineering~Functionality}

\keywords{LLM, Design Pattern, Program Synthesis}

\maketitle

\section{Introduction}
Large Language Models (LLMs) have gained acceptance in the software engineering industry thanks to tools like GitHub Copilot and Claude Code. They provide software engineers with a boost in productivity in implementation, testing, requirements engineering, and code review. At the same time, software engineering is more than just programming or testing; it is also about good design, robust software construction, and reliable software operations. Software design patterns are a good example of software engineering knowledge that is used in many industries to standardize the implementation of recurring patterns in software \cite{gamma1995design}. It is a way for developers to use a shared language among themselves and to instantly identify the behavior of design concepts \cite{buschmann2007pattern}. Architectural patterns and secure software design patterns are additional examples of these \cite{nguyen2015sospa}.  

In the age of LLMs, an ever-increasing share of code is generated by these models, and they have become an essential tool for many developers by assisting with tasks ranging from implementing new features to refactoring \cite{SERGEYUK2025107610}. However, while these models can generate functional, syntactically correct code at the low level, they do not adhere to higher-level design principles such as modularity and abstraction \cite{GPTsurvey}. In particular, design patterns are seldom applied correctly in LLM-generated code \cite{DoLLMunderstandDP}. 

The authors of the classical GoF (Gang of Four) patterns \cite{gamma1995design} have identified 23 common design patterns, which are studied extensively in software engineering. One of these is the Singleton design pattern, whose intent is defined as to \textit{"Ensure a class only has one instance, and provide a global point of access to it"} \cite{gamma1995design}. This pattern is used in complex systems where it is important that only a single instance of a given class can exist at any given time. This could be logging, database connections, or the software object representing a car engine. The Singleton design pattern is a creational pattern; its properties are defined during code creation, making it easy to recognize and automatically detect.

In this paper, we investigate how to guide LLMs towards generating code that follows the Singleton design pattern. This study is important to us because it helps us determine whether we can use existing LLMs alongside design pattern recognition/design software to generate new software code. We conduct a computational experiment where we explore prompting strategies and tool-assisted automated feedback strategies. We also study whether these strategies have an impact (potentially detrimental) on the functionality of the generated source code.

Firstly, we study how to guide the LLMs to generate code that follows the Singleton design pattern using direct instructions. 
 
\emph{RQ1: To what extent do direct instructions prompting strategies influence a model’s ability to generate codes that correctly implement the Singleton design pattern?}

In this research question, we evaluate whether models can successfully generate code that follows the Singleton design pattern when instructed to do so.

Secondly, we provide the models with feedback on how well the generated code aligns with the Singleton pattern and allow them additional attempts to generate correct code. 

\emph{RQ2: To what extent does automated feedback help LLMs align the code to the Singleton design pattern?}

We experimented with binary feedback, where we solely inform the LLM whether the generated code follows the Singleton design pattern, as well as with more extensive feedback indicating which properties of Singleton are fulfilled and which are not. The extensive feedback was also paired with examples of correctly implemented Singleton classes to help the LLMs. 

Lastly, we want to understand how these strategies will affect our ability to implement the Singleton pattern without negatively affecting the overall functional correctness of the generated code. This is formulated as the following research question:  

\emph{RQ3: To what extent can LLMs be guided to generate Java classes following the Singleton design pattern, without negatively affecting the functionality of the code?}

Our experiments show that we can guide LLMs to implement the Singleton design pattern in the generated code without negatively affecting their functionality. However, the optimal strategy depends on which LLM is used. 

The rest of the paper is structured as follows. Section \ref{sec:related_work} reviews previous studies related to design patterns and prompting strategies. In Section \ref{sec:method}, the dataset, evaluation metric, and requirements for a Singleton class are presented. Section \ref{sec:experiments} consists of a detailed description of the experiment design used in the four experiments. The results of the trials are presented in \ref{sec:results}. 


\begin{figure*}[t]
\centering
  \includegraphics[width=14cm]{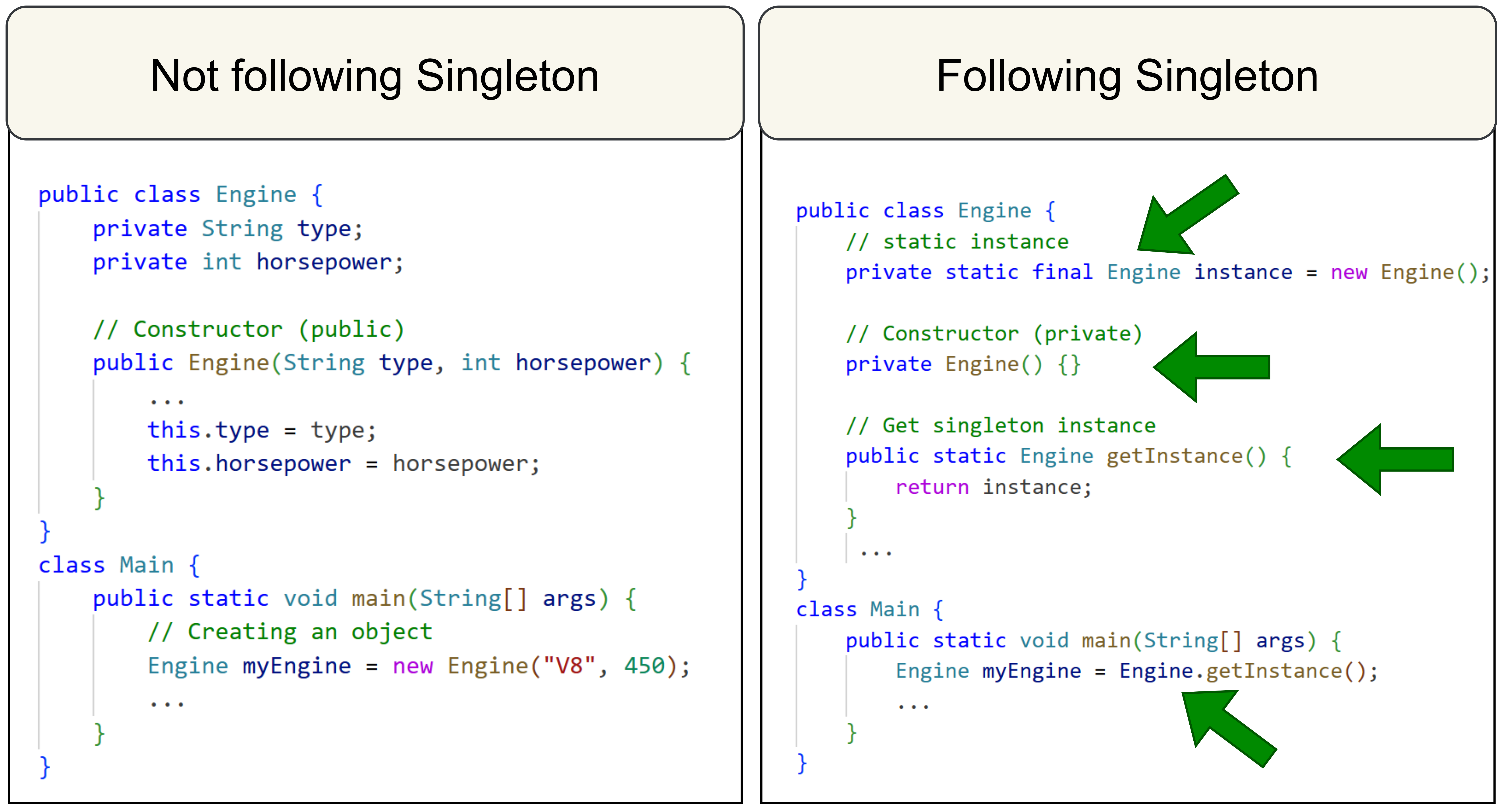} 
  \caption{A class representing the software of an engine. To the left, without implementing Singleton. On the right, the engine class is implemented as a Singleton. The arrows point toward the rows, which show that it is a Singleton class.}
  \label{fig:intro_ex}
\end{figure*}
\section{Related Work}
\label{sec:related_work}
Our study builds on the existing work in design pattern recognition, automated design, and prompting strategies in software development. 


\subsection{Automated Design}

\citet{tian2025fixing} uses a method called \(\mu\)FiX to refine the program's specifications in the tasks. The system receives a natural-language explanation of the expected output along with a definition of the function. The LLM then generates a natural-language description of the specifications and iteratively identifies logical errors in the specification until it generates a correct specification. The specification is then enriched, and an initial code is generated from it. The initial code is then tested against predefined test cases and adjusted until it passes them. \(\mu\)FiX achieves a higher Pass@1 than 9 other state-of-the-art prompting techniques, including Few-shot prompting and CoT (Chain-of-Thought) prompting. The study evaluates only the functionality of the generated code and does not consider higher-level design or architectural patterns.

\subsection{LLMs and Design Patterns}

\citet{TransDPR} presented a pre-trained TransCoder-based model that extracted semantic context from source code, which was then classified into design patterns using a logistic regression layer. TransDPR, as the model was called, was then evaluated on both open-source and industry projects and achieved 90\% accuracy in categorizing different design patterns. Our study expanded on that work by introducing the ability to generate design patterns, not only detecting them.

\citet{DoLLMunderstandDP} evaluated the capabilities of LLMs in defining, completing lines, and generating functions based on design patterns. They showed that the models struggled to classify design patterns and that Singleton patterns are often over-predicted and misclassified because of their structures, which overlap with those of other patterns. When generating functions for classes of different design patterns, the models achieved low code similarity, even with knowledge of the design pattern of the given class. In some cases, knowledge of the design pattern negatively affected performance, indicating the need for improved contextual knowledge over the related design patterns. We expanded on their study by introducing multiple prompting strategies and attempting to fill the identified gap in contextual knowledge through automated feedback on the implementation of Singleton in the generated code.

\citet{KIM2025112519} compared the ability of ChatGPT, Claude, Copilot, Gemini, and Meta on detecting design pattern applicability in source code and implementing applicable design patterns. They assessed these models' capabilities by an evaluation framework using three metrics: the Property Satisfactory Rate, Critical Property Coverage, and Pattern Implementation Quality Score. The results varied across models, with Claude achieving the highest Pattern Implementation Quality Score. They used a straightforward prompting strategy, asking each model to apply the chosen design pattern in a specific location. However, they suggest that, while LLMs have potential, they are best used as tools to assist human developers rather than as replacements. They argued that the results indicated that incorporating contextual knowledge could increase the model's performance. They evaluated multiple models with multiple design patterns; however, the limitation was that they only used direct instructions as the strategy of guiding the LLMs to identify and generate according to the design pattern. Here we expand on their study by evaluating multiple strategies. As well as by using a larger dataset rather than two small cases, in which Singleton was implemented in only one class per test case.

\subsection{Self-improvement}
Recent advances in LLMs have increased their ability to generate cohesive, increasingly strong performance. However, their performance remains highly correlated with the contextual information they have access to: the same model can produce different results for the same task depending on the prompting strategy used.

In \citet{selfRefined}, they used a multi-iteration self-refine algorithm that iteratively refines the initially generated answer through an iterative process of generating feedback, which was used to improve the quality of the final generated answer. Self-refinement consistently improved results across a variety of base models, sizes, and tasks. It could be used to improve translation \cite{chen2024iterative}, code synthesis \cite{NEURIPS2023_1b44b878}, and computer tasks \cite{NEURIPS2023_7cc1005e}, to name a few. Self-refinement is usually performed in a multi-iteration setting where the highest gains are achieved in the first three iterations. We leveraged self-refinement and the iterative process across three of our experiments.

An alternative to relying on the model's ability to reflect on how to improve its own generated answers is to combine LLMs with tools. These tools can be code or test executors, compilers, or APIs for a range of applications. They can be an integrated module in the process, as each program generated by the LLMs is executed \cite{chen2023teaching} or as a test executor who executes the related unit-test after and uses the test result as feedback \cite{peng2025perfcodegen}. We leveraged tools that checked the rate of Singleton integration in our experiments to boost LLM performance and implement Singleton in the generated code.

Providing code examples to the LLM via the input prompt can help steer it towards the desired coding style with minimal added complexity \cite{NEURIPS2020_1457c0d6}. Recent studies show that the few-shot learning (also known as in-context learning) effect depends on how examples or demonstrations are constructed and in what order they are presented \cite{10298329}. We took advantage of this by experimenting with providing examples alongside informative feedback to the models to achieve a higher rate of Singleton implementations in the generated code.

\section{Method}
\label{sec:method}

\subsection{Experiment design}

\begin{figure*}[t]
\centering
  \includegraphics[width=18cm]{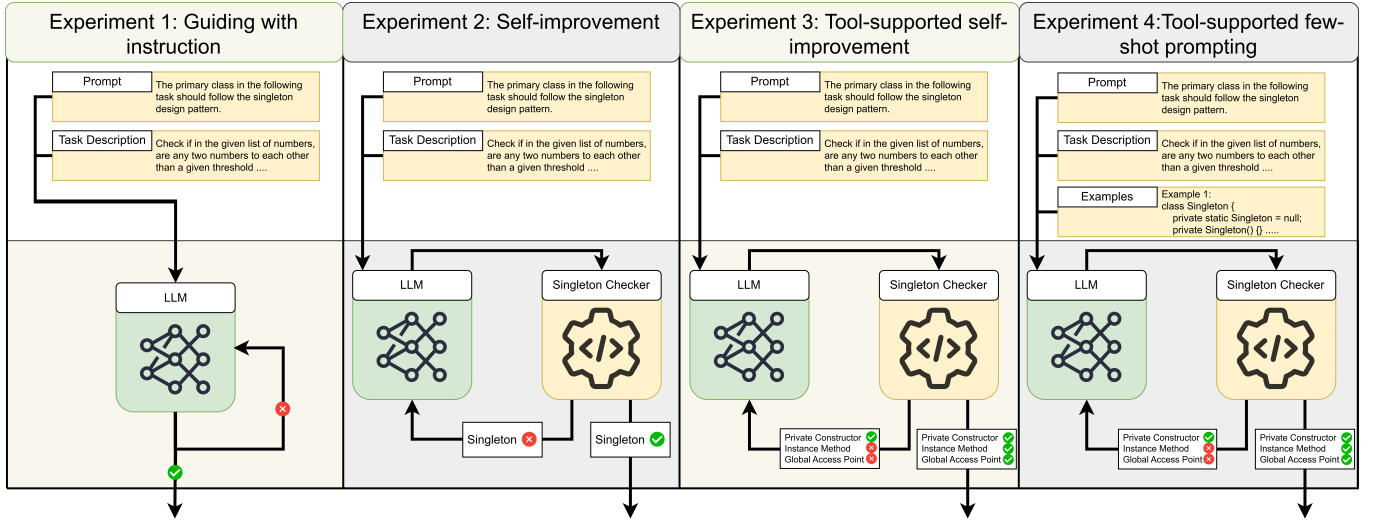} 
  \caption{The design and process of the four experiments. In each experiment design the input is shown in the white boxes with the LLM and potential interaction with the Singleton Checker below.}
  \label{fig:experimentSetup}
\end{figure*}

The study consisted of four computational experiments in which different strategies were used to guide the LLMs to implement the Singleton design pattern into the generated code. The four strategies were called \textbf{guiding with instructions}, \textbf{self-improvement}, \textbf{tool-supported self-improvement}, and \textbf{tool-supported few-shot prompting}. 
The same dataset was used in all four experiments, and the results were evaluated with the same metrics. Each of the experiments is described in more detail in Section \ref{sec:experiments}, but an overview of the design can be seen in Figure \ref{fig:experimentSetup}.

\subsection{Dataset}

The first criterion for the dataset was that it should consist of tasks that did not have canonical solutions requiring Singleton, to evaluate how to guide LLMs to generate code implemented with Singleton. The second criterion was that the dataset should be in an object-oriented language. Thirdly, there should be test cases for each task in the data set to verify the functional correctness of the generated code. 

The chosen dataset, which met all requirements, was HumanEval-X\footnote{https://huggingface.co/datasets/zai-org/humaneval-x}, consisting of 164 problems with task descriptions and test cases. The dataset does not consist of problems where the canonical solution is computed with a Singleton design pattern, and we needed such a dataset to ensure that no solution was "by accident" a singleton, which would bias our results. We used each task description as a base and generated solutions using the Singleton pattern. Each generated solution can be tested using the predefined test cases for each task in the dataset.

\subsection{Functionality}
The LLMs generated code that followed the Singleton design pattern; however, adapting to this design pattern should not affect the code's overall functionality. Here we define functionality as how well the generated code fulfilled the requirements on functionality. In other words, if the generated code passed the defined test cases for the given task. Each model was only given a single attempt to generate a candidate solution using each strategy. Therefore, we evaluated the generated code against the \textbf{pass@1} rate on the test cases in the dataset. Here, pass@1 indicate that we only allowed one attempt to generate a solution that passed the tests for each iteration for each model-strategy pair. With this, we could assess the effect of the Singleton design pattern requirement on the code's functionality by calculating the number of solutions that pass the related tests for each task.

\subsection{Singleton}
\label{sec:req_singleton}
To illustrate how Singleton can be used, imagine a car engine represented as a class in a project to develop a new car, where the class manages all variables related to the engine (thrust, clutch, fuel flow, etc.). When the rest of the system running the car needs to interact with any of these functions, they should do so through the same instance of the engine class. The Singleton design pattern ensure that this is done correctly.
The difference between an engine class implemented with and without Singleton can be seen in Figure \ref{fig:intro_ex}.

Singleton is a creational design pattern with identifiable predicates, and mainly affects how a specific class is created and imposes no limitations on its behavior, except that there can only be a single instance of the class \cite{singleton_def}. 
Instead of creating multiple instances of the object, the single object of a class following the Singleton design pattern should be reused. The exact design of a Singleton class can vary between different implementations; however, the canonical definition of the intent of a Singleton class can be said as follows: \textit{Ensure a class only has one instance, and provide a global point of access to it.} \cite{gamma1995design}. 

The intent of the Singleton class can be broken down into specific properties that must be present inside the class. In a Singleton class, all the constructors should be private to prevent external instantiation. There should also be a static instantiation method present that returns the instance of the singleton. The class also needs to be globally accessible. These properties can be used to assess how closely a class adheres to the Singleton design pattern. Based on the intent, we can define predicates for the class that a given class follows the Singleton design pattern:

\begin{description}
  \item[$\bullet$ Private Constructor] The class has a single constructor method, which is private.
  \item[$\bullet$ Instance Method] There is a private static method for retrieving the object.
  \item[$\bullet$ Global Access Point] The instance should be globally retrievable.
\end{description}

Based on the Pattern Property Satisfactory Rate (PSR) described in \citet{KIM2025112519}, the alignment towards Singleton could be evaluated for each generated solution. The PSR measures how well a class adheres to a specific design pattern by the number of properties it satisfies. However, Singleton only has a single property (whether it is a Singleton or not); therefore, a modified version of PSR was used, in which the number of predicates in the class was measured instead of the number of properties. The measurement, called the Singleton score, was defined as the following:    
\begin{center} 

$ Singleton Score = \frac{\#missing\_predicates}{\#total\_predicates} * 100 $
\end{center}

Here, the Singleton score can be calculated for each instance generated by the LLMs in the dataset. The average Singleton score can then be calculated based on the score of all generated instances of an LLM. This gives a score indicating the average rate of Singleton implementation for each model. 

The check of which predicates were fulfilled in the generated class was done through a number of regular expression statements in an automatic program. First, Private Constructor was checked by ensuring that the class had a private constructor and no public constructors. Second, Instance Method was set to true if the class included a private and static instance field. Lastly, Global Access Point was set to true if the class included a public static accessor method.

\subsection{McNemar's test}
We used McNemar's test \cite{mcnemar1947note} to statistically confirm whether the differences in the test case pass rates across experiments were significant. It is used to compare two methods on the same data set of subjects with a binary outcome and, in the context of this study, to compare the pass/fail rate of each experiment with the baseline. A benefit of using McNemar's test is that it compares binary results pairwise and can therefore detect differences in pass/fail outcomes for each test.

\section{Experiments}
\label{sec:experiments}
Thirteen models were selected for all experiments, including both open-weights and commercially available models. They included LLMs from six different families of models: GPT, Llama, Qwen, Gemma, Mistral, and DeepSeek. They were in a variety of sizes and included both general-purpose and code-specific models. All model families were chosen based on their proven abilities in common use Software Engineering research, as seen in the references in Table \ref{tab:models}.

\begin{table}[t]
\centering
\scriptsize
\setlength{\tabcolsep}{5pt} 
\renewcommand{\arraystretch}{1.1}
\caption{The chosen models with the size, access status, role and references to studies that study or report their use in Software Engineering}
\label{tab:models}
\begin{threeparttable}
\begin{tabularx}{\columnwidth}{@{}l r c c l@{}}
\toprule
\textbf{Model} & \textbf{Sizes} & \textbf{Access} & \textbf{Role} & \textbf{Reference} \\
\midrule

GPT-5 mini & -- & Closed & Generalist & \cite{wang2026codeflowbenchmultiturniterativebenchmark} \\
GPT-4o mini & -- & Closed & Generalist & \cite{wang2026codeflowbenchmultiturniterativebenchmark} \\
GPT-OSS & 20B & Open & Generalist & \cite{bi2025gptossgoodcomprehensiveevaluation} \\
Llama 3.3 & 70B & Open & Generalist & \cite{DoLLMunderstandDP, bi2025gptossgoodcomprehensiveevaluation} \\
Code Llama & 70B & Open & Coder & \cite{10952968} \\
Gemma 3 & 4B, 27B & Open & Generalist & \cite{ bi2025gptossgoodcomprehensiveevaluation} \\
Qwen3 & 8B, 32B & Open & Generalist & \cite{yang2025qwen3technicalreport} \\
Qwen3 Coder & 30B & Open & Coder & \cite{DoLLMunderstandDP} \\
Mistral & 7B & Open & Generalist & \cite{Samo_Ali_Memon_Abbasi_Koondhar_Dahri_2024} \\
DeepSeek R1 & 70B & Open & Reasoning & \cite{10952968,bi2025gptossgoodcomprehensiveevaluation} \\
DeepSeek Coder v2 & 16B & Open & Coder & \cite{DoLLMunderstandDP, 10952968} \\
\bottomrule
\end{tabularx}
\end{threeparttable}
\end{table}

\subsection{Baseline for functionality}
\label{sec:baseline}
Each LLM was given the task definitions and prompted to generate the appropriate code for each. The models were not instructed to generate code following any specific design pattern, but instead they were instructed through the role description to give a solution written in Java without comments and explanation: 

\textit{You are a Java programmer. You respond with the code in Java to solve the task. No comments or explanations}

This description of the role was consistent across all models and experiments.
The results served as a baseline for the functionality in the following experiments, in which we used instructions and automatic feedback to guide the LLMs to generate classes implementing the Singleton pattern.  

\subsection{Experiment 1: Guiding with instruction}

Here, we evaluated how likely the models were to generate code that followed the Singleton design pattern when instructed directly to do so. 
The models were given the initial class declaration and the task description. 
In addition to the role description (see Section \ref{sec:baseline} ), we provided the models with the following prompt: \textit{The primary class in the following task should follow the singleton design pattern.\{task description\}}.

Each model was given the empty class definition from the data set and the description of the natural-language problem. 
The models were not provided with a definition of a Singleton class or acceptance requirements. 
Instead, the LLMs had to rely on their own weights gained through training. 

The models were then given 10 iterations to generate a correct Singleton class, and the first, if any, correct implementation was chosen as their answer. 
The models were not given feedback on which properties were missing in the generated candidate code, but simply given the following prompt after each iteration: \textit{Make sure that the primary class in the following code follows the singleton design pattern. \{candidate code\}}. 

The hypothesis was that the models would be expected to follow the instruction and generate classes that fulfill some of the requirements, but, without access to the definition of a Singleton class, would be unable to meet all requirements.
We evaluate the generated code for functionality and the Singleton Score.

\begin{table*}[htbp]
    \centering
    \footnotesize
    \setlength{\tabcolsep}{8pt} 
    \caption{Increased rate of passed test cases in percent for each model across prompting setups. The table is sorted after the highest pass rate for the baseline. McNemar significance indicated with: * = p<0.05, ** = p<0.01, *** = p<0.001}
    \label{tab:test-rate-setups}
    \begin{tabular}{llllll}
        \toprule
        Model &
        Baseline (\%) &
        \shortstack{Guiding with\\instructions} &
        \shortstack{Self-\\improvement} &
        \shortstack{Tool-supported\\self-improvement} &
        \shortstack{Tool-supported\\few-shot\\prompting} \\
        \midrule
        GPT-5 mini               & 83.5 & -1.2 & +0.0 & -0.6 & +0.0 \\
        GPT-OSS (20B)           & 75.6 & +6.1* & +6.1* & +6.1* & +4.3 \\
        Gemma3 (27B)            & 72.0 & +0.0 & +0.6 & -1.2 & -1.2 \\
        DeepSeek Coder v2 (16B) & 61.0 & -26.2*** & -9.8** & -11.0** & -17.1*** \\
        Gemma3 (4B)             & 53.7 & -21.3*** & -21.3*** & -22.0*** & -32.3*** \\
        DeepSeek r1 (70B)       & 51.8 & -7.9 & -3.7 & +6.7 & +13.4** \\
        Qwen3 (32B)             & 49.4 & +27.4*** & +27.4*** & +23.8*** & +22.6*** \\
        Qwen3 Coder (30B)       & 42.7 & +32.9*** & +31.1*** & +31.1*** & +29.3*** \\
        Qwen3 (8B)              & 40.9 & +8.5 & +17.7*** & +12.2 & -8.5 \\
        Llama3.3 (70B)          & 32.9 & +34.1*** & +28.7*** & +31.7*** & +14.0** \\
        Code Llama (70B)        & 18.9 & -9.1* & -6.1 & -7.3 & -12.8*** \\
        Mistral (7B)            & 11.0 & -1.2 & +6.1 & +2.4 & -3.7 \\
        GPT-4o Mini              & 4.3  & +55.5*** & +61.0*** & +64.0*** & +22.0*** \\
        \bottomrule
    \end{tabular}
\end{table*}

\begin{table*}[htbp]
    \centering
    \footnotesize
    \setlength{\tabcolsep}{8pt} 
    \caption{Average SingletonScore for each model in different prompting strategies. Sorted by the best performance from the last experiment}
    \label{tab:singleton-scores}
    \begin{tabular}{lrrrrr}
        \toprule
        Model &
        Baseline &
        \shortstack{Guiding with\\instructions} &
        \shortstack{Self-\\improvement} &
        \shortstack{Tool-supported\\self-improvement} &
        \shortstack{Tool-supported\\few-shot\\prompting} \\
        \midrule
        Llama3.3 (70B)          & 0.0 & 100 & 100.0 & 100.0 & 100.0 \\
        GPT-4o mini              & 0.0 & 99.8 & 100.0 & 100.0 & 100.0 \\
        GPT-OSS (20B)           & 0.0 & 99.0 & 100.0 & 100.0 & 100.0 \\
        GPT-5 mini               & 0.0 & 98.6 & 100.0 & 100.0 & 100.0 \\
        Gemma3 (27B)            & 0.0 & 100 & 100.0 & 100.0 & 100.0 \\
        Qwen3 Coder (30B)       & 0.0 & 98.8 & 99.2 & 100.0 & 100.0 \\
        Qwen3 (32B)             & 0.0 & 98.8 & 99.6 & 100.0 & 99.4 \\
        Qwen3 (8B)              & 0.0 & 98.8 & 99.2 & 99.4 & 95.7 \\
        Mistral (7B)            & 0.0 & 91.8 & 95.3 & 98.4 & 95.3 \\
        DeepSeek r1 (70B)       & 0.0 & 32.9  & 72.2 & 73.2 & 85.4 \\
        DeepSeek Coder v2 (16B) & 0.0 & 34.1 & 54.5 & 80.1 & 71.3 \\
        Code Llama (70B)        & 0.0 & 14.2 & 29.5 & 36.4 & 43.3 \\
        Gemma3 (4B)             & 0.0 & 87.2  & 78.1 & 82.9 & 40.7 \\
        \bottomrule
    \end{tabular}

\end{table*}

\subsection{Experiment 2: Self-improvement}
In this experiment, we evaluated LLMs' ability to generate code in accordance with the Singleton design pattern by providing binary feedback if the code includes a Singleton class consisting of all three predicates explained in section \ref{sec:req_singleton}. 
For each model, the same coding tasks were provided as in the previous experiment. 
After each attempt, the solution candidate was automatically evaluated to verify whether it met the Singleton design pattern's structural requirements. 
If the code did not include all the properties of a Singleton class, the following feedback was given to the model:

\textit{The following code does not include a correctly formatted Singleton class: \{response\_singleton\}. Please correct the code and return the complete code.}

In this experiment, the models were not informed which properties were fulfilled and which were not; they were only informed that the generated code did not conform to the Singleton design pattern. The model was then given another attempt to generate code aligning with the Singleton design pattern.

The purpose of this experiment was to analyze whether the models could correct the generated code solely by being informed of whether the requirements were fulfilled. If the models were trained on enough Singleton examples, they could have learned to recognize the design pattern and recognize what element is missing from the previous attempt. The hypothesis was that we would see an increase in the number of correctly structured classes following Singleton, but that the model's size would affect how well they could internalize the feedback and correct the code accordingly. 

\subsection{Experiment 3: Tool-supported self-improvement}
Instead of only giving binary feedback on the correct usage of the pattern, we provide feedback on which Singleton properties are not present in the generated code.

Here, the same setup as in the previous experiment, with a feedback loop for the LLMs on whether the generated candidate followed the Singleton design pattern, but with enriched feedback. As before, each model was given up to ten iterations to generate a correct Singleton implementation, and after each attempt, the candidate was analyzed. However, instead of simply returning a binary indicator, the results of each check were reported to the LLMs. The LLMs were therefore informed exactly which properties were missing from the generated candidate. This informative feedback enabled the LLMs to adjust their candidate implementation based on explicit indications of missing properties, enabling us to study how explicit guidance influences their ability to converge to a correct Singleton implementation. The prompt given to each LLM, including the feedback, was as follows:

\textit{The following code does not include a correctly formatted Singleton class: \{response\_singleton\}. It failed the following checks \{singleton\_errors\}. Please correct the code and return the complete code}

For this experiment, we hypothesized that informing the models of which requirements were not met would enable them to adjust the candidate solution to better align with the requirements for a class implementing the Singleton Design Pattern.

\subsection{Experiment 4: Tool-supported few-shot prompting}

The last experiment executed was with automated feedback on alignment to Singleton, along with two examples of correctly implemented Singleton classes. The models then had access to the requirements that were fulfilled and to examples of two classes that fulfill them. 

The hypothesis was that with providing additional examples of correctly structured Singleton classes, the models would generate solutions with the Singleton design pattern to a higher degree and help low-scoring models.

\section{Results}
\label{sec:results}

\begin{figure*}[htbp]
    \centering

    \begin{subfigure}[b]{0.48\textwidth}
        \centering
        \includegraphics[width=\textwidth,trim=12 14 11 11,clip]{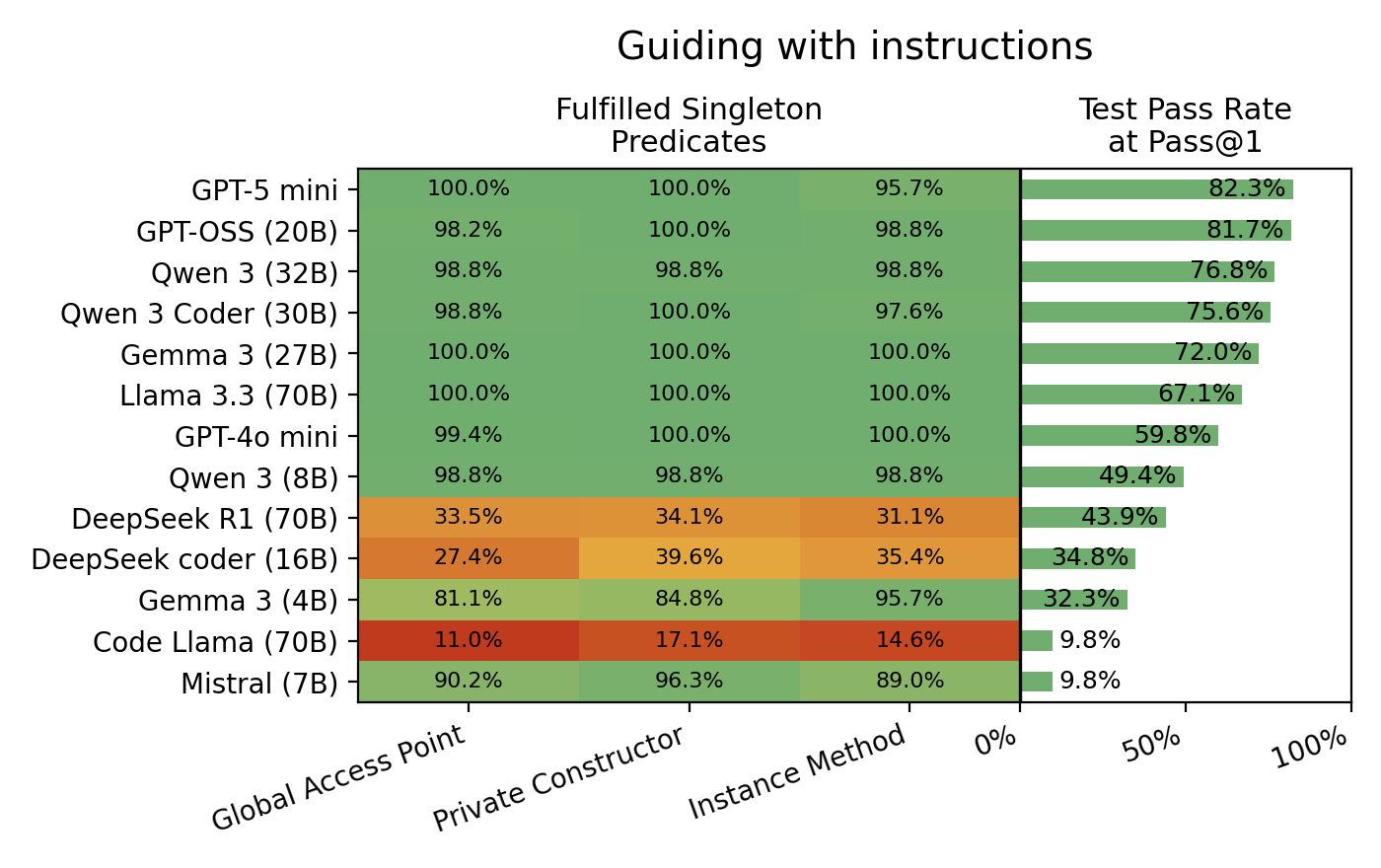}
        \caption{Experiment 1: Guiding with instructions}
        \label{fig:guiding}
    \end{subfigure}
    \hfill
    \begin{subfigure}[b]{0.48\textwidth}
        \centering
        \includegraphics[width=\textwidth,trim=12 14 11 11,clip]{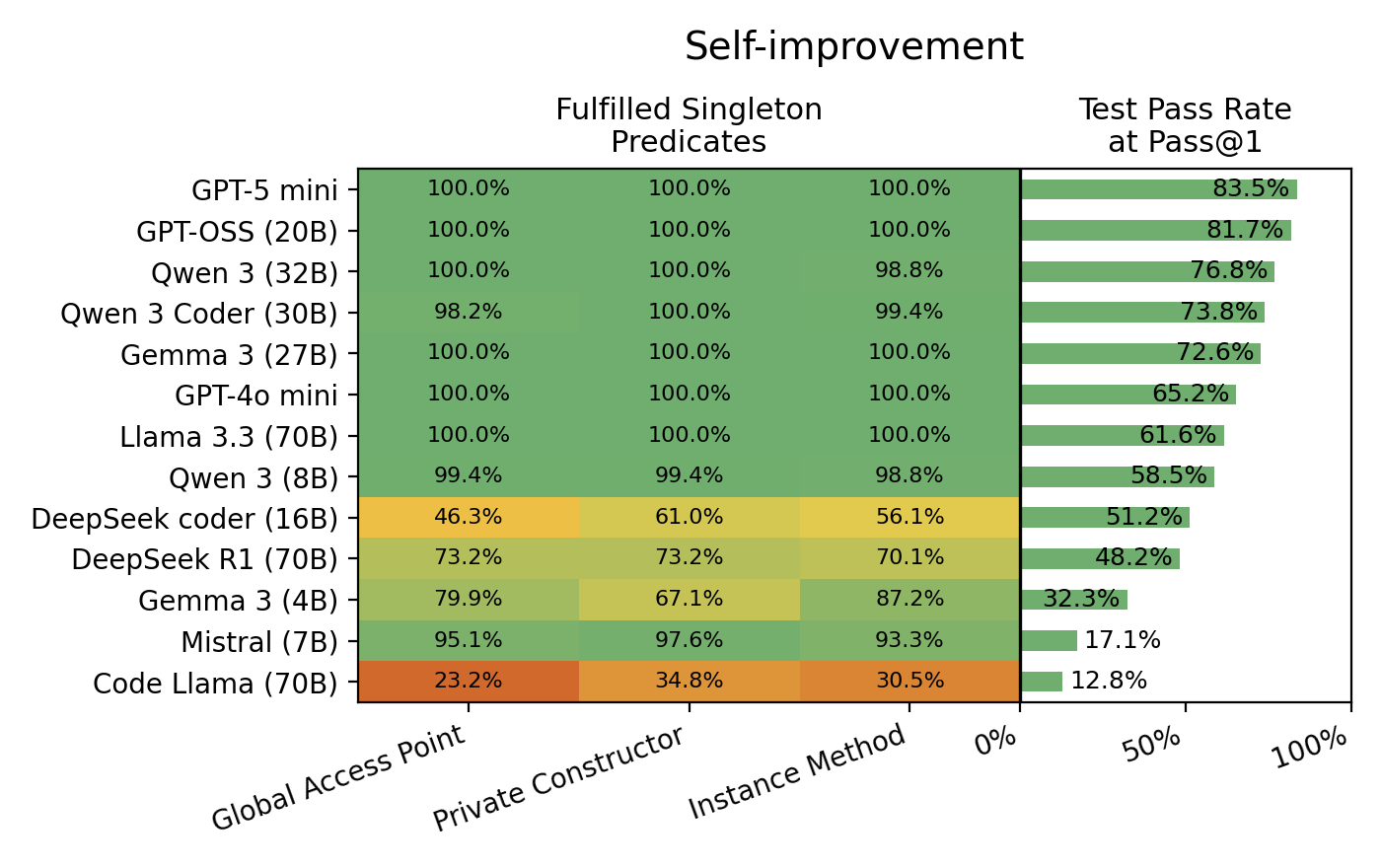}
        \caption{Experiment 2: Self-improvement}
        \label{fig:self_imp}
    \end{subfigure}

    \vspace{0.5cm}

    \begin{subfigure}[b]{0.48\textwidth}
        \centering
        \includegraphics[width=\textwidth,trim=12 14 11 11,clip]{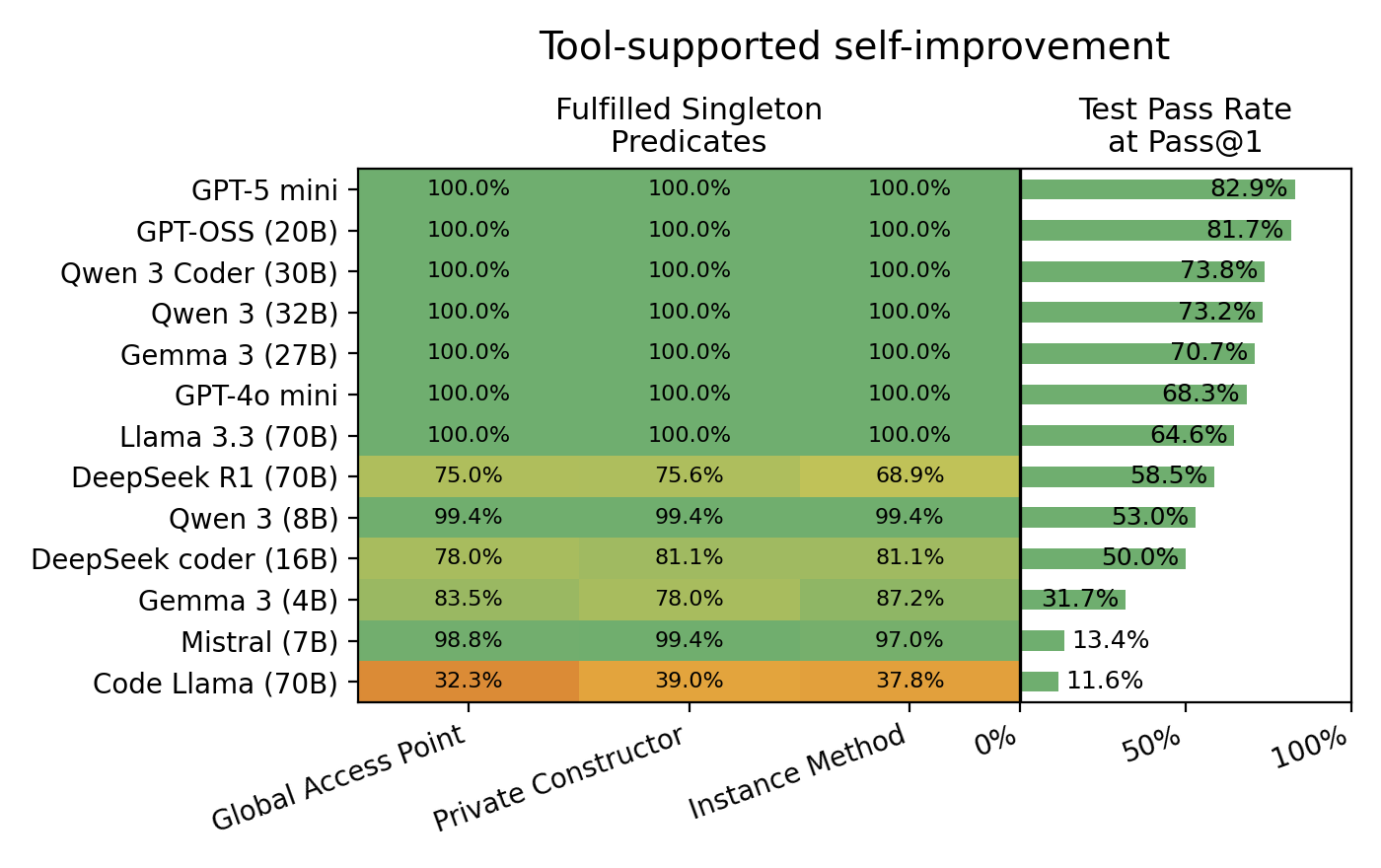}
        \caption{Experiment 3: Tool-supported self-improvement}
        \label{fig:tool_self_imp}
    \end{subfigure}
    \hfill
    \begin{subfigure}[b]{0.48\textwidth}
        \centering
        \includegraphics[width=\textwidth,trim=12 14 11 11,clip]{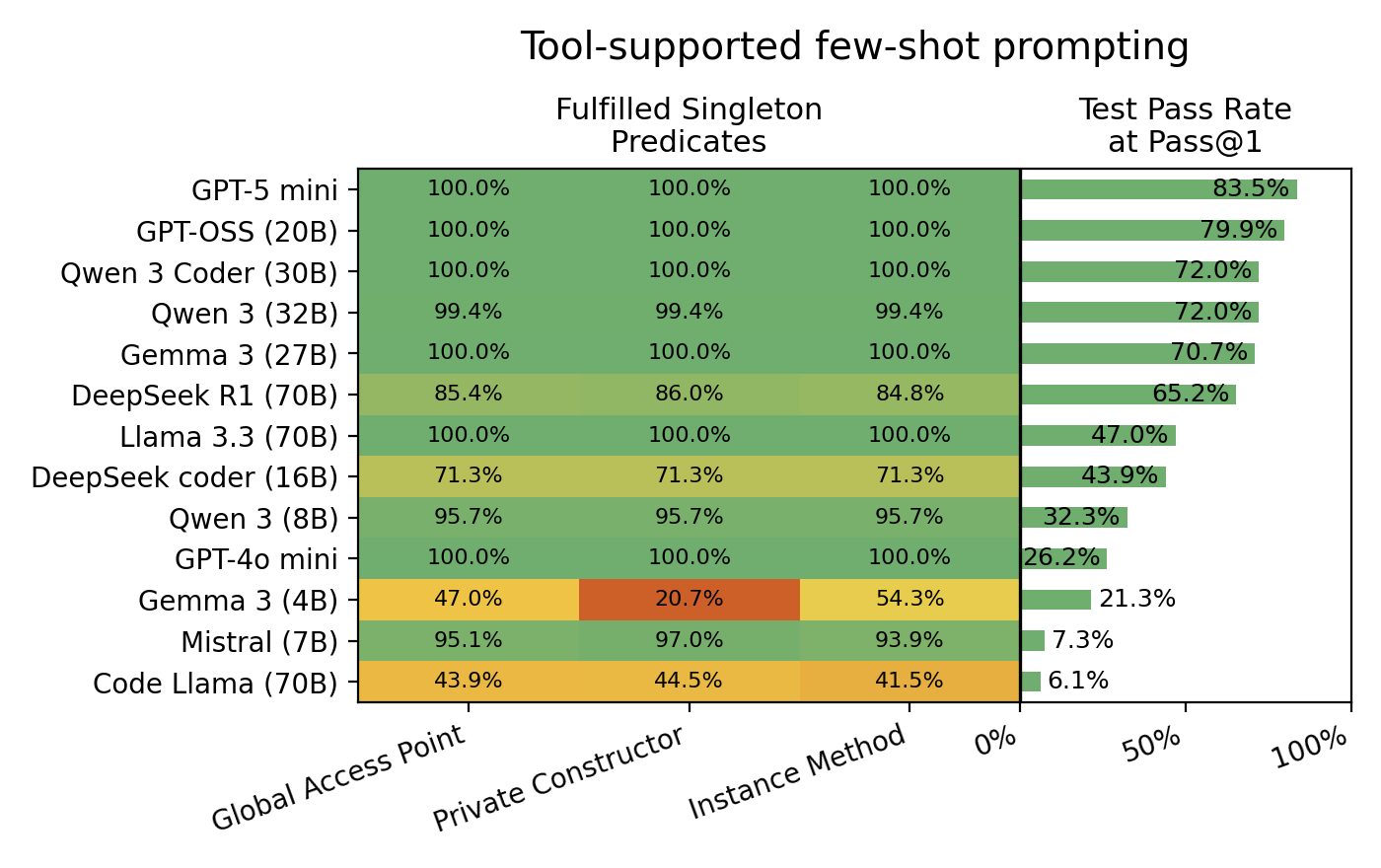}
        \caption{Experiment 4: Tool-supported few-shot prompting}
        \label{fig:tool_few_shot}
    \end{subfigure}

    \caption{On the left side of each sub-figure: the number of fulfilled predicates per model. On the right side of each sub-figure: the rate of passed test, pass@1 for each model.}
    \label{fig:four_panel}
\end{figure*}

\subsection{Functionality Baseline}
The models were asked to generate solutions to the problems in HumanEval-X and were evaluated based on whether they passed the tests for each task. This provided a way to compare how the functionality of each class was affected by the additional requirement to incorporate the solution into a class that followed the Singleton design pattern.

In Table \ref{tab:test-rate-setups}, the percentage of solutions with tests passed for each model is shown. The results vary between models; however, none generated solutions that passed all the tests. GPT-5-mini achieved the highest test pass rate of 83.5\%, followed by GPT-OSS and Gemma3 (27B). 

The average Singleton Score indicated that none of the models generated classes that conformed to the Singleton design pattern (see Table \ref{tab:singleton-scores}). However, the metric was still used to measure whether any models would spontaneously generate solutions following this design pattern for the given problems.

\subsection{Guiding with instruction}
\label{sec:ex1}

\subsubsection{Singleton Score (Table \ref{tab:singleton-scores})}
All models increased their rate of successful Singleton implementation when instructed to generate solutions that conform to the Singleton design pattern; however, the magnitude of this effect varied across models. 
Several models successfully generated solutions with the correct properties and achieved an average Singleton Score above 90. In particular, Llama3.3 and Gemma3 (27B) both achieved a perfect Singleton Score of 100.

\subsubsection{Rate of Passed Test Cases (Table \ref{tab:test-rate-setups})} When it comes to the change in test pass rate from the Baselines, each model was affected differently.
Some models, such as DeepSeek-Coder and Gemma3 (4B), exhibit a significant adverse effect: the number of solutions that passed the tests was lower for those instructed to follow the Singleton pattern. Some models had the opposite effect; Llama3.3, GPT-4o Mini, and Qwen3-Coder were among those in which this strategy led to more solutions passing the tests. The increase for all these three models was significant as indicated in Table \ref{tab:test-rate-setups}. So, even with the additional requirement of following the Singleton design pattern, they generated solutions of higher quality than before. GPT-4o Mini increased the pass rate significantly by 55.5 percentage points compared to the Baseline.   

\subsubsection{Fulfilled Singleton Predicates and Pass Rate (Figure \ref{fig:guiding})} 
The general pattern here is that the models that achieved a high pass rate also fulfilled all predicates to a great extent. Many of the LLMs were able to fulfill the predicates to about the same degree across all three predicates. 

Mistral fulfilled all three predicates to a high degree; however, it achieved a test pass rate of only 9.8\%. A majority of the generated solutions contained compiler errors, as shown in Figure \ref{fig:pass_rate}. A manual evaluation of the generated solutions was conducted, and 55\% of the compiler errors were caused by references to external libraries not included.  

The other model that stuck out was Code Llama, which was one of the largest models included but still performed the worst in both fulfilled predicates and the test passed rate. When the generated result from this model was examined, 76\% of all compiler errors were caused by the model failing to generate code and instead only generating textual output. This can also explain the low rate of fulfilled predicates.


\begin{tcolorbox}[colback=beige, colframe=beige, boxrule=0pt,
                  left=8pt,right=8pt,top=8pt,bottom=8pt, breakable]
\textbf{Answer to RQ1:} Direct instruction has a clear and generally positive effect on the model's ability to generate solutions following the Singleton design pattern, often dramatically improving the model's ability and enabling almost perfect implementations for some models. However, the influence is highly model-dependent and can lead to a trade-off against the functional correctness. 
\end{tcolorbox}

\subsection{Self-improvement}
\subsubsection{Singleton Score (Table \ref{tab:singleton-scores})}
When models are allowed to iteratively improve the generated code until it satisfies the Singleton design pattern requirements, the rate of code with complete Singleton implementations is higher compared to the Baseline . Multiple models (Llama3.3, Gemma3 (27B,), GPT-OSS, GPT-4o Mini and GPT-5 Mini) generated solutions following the Singleton design pattern for all given tasks.
All three Qwen3-based models also performed well, with a Singleton implementation rate above 97\%. 

\subsubsection{Fulfilled Singleton Predicates and Pass Rate (Figure \ref{fig:self_imp})} 
Some models showed an increase in the number of successfully implemented Singleton classes compared to Guiding with instructions. Still, Code Llama and DeepSeek Coder achieved the worst results, with rates of 29.5\% and 54.5\%, respectively. Both models generated solutions with missing predicates; the global access point was the predicate most often missing in the generated solutions.  

This strategy helped several models, in which a majority of the LLMs generated solutions with all predicates nearly perfectly fulfilled. 
DeepSeek Coder, DeepSeek, and Code Llama saw an increase in successfully implemented Singleton classes compared to previous experiment.

\subsubsection{Rate of Passed Test Cases (Table \ref{tab:test-rate-setups})}
Not only did the rate of successfully implemented Singleton classes increase compared to the baseline, but the rate of passed tests also increased for seven of the models. Llama 3.3 increased the pass rate by 28.7 percentage points to 60.5\% on the tests. However, DeepSeek Coder and Gemma3 (4B) showed worse results than the baseline, with decreases of 9.8 and 21.3 percentage points, respectively, significantly worse results than the baseline. For DeepSeek Coder, the rate of implemented Singleton cases increased, indicating that the model generated solutions that followed Singleton, but did not meet the functionality requirements. This means that the model "lost" the context and implemented the pattern correctly, but for the wrong functionality. 

\subsection{Tool-supported self-improvement}
\subsubsection{Singleton Score (Table \ref{tab:singleton-scores})}
By giving the models feedback on which predicates were missing from the generated code, the number of models that were able to implement the Singleton design pattern into all generated code was increased to seven in comparison to the five as was the case in the previous experiment:  Gemma3 (27B), GPT-OSS, Llama3.3, Qwen3 Coder, and Qwen3 (32B), GPT-4o Mini and GPT-5 Mini. Two more models, Qwen3 (8B) and Mistral (7B), achieved a score above 97.0\%. However, all models were able to increase the rate of implementing the Singleton design pattern in the generated problems, or achieve a similar rate, compared to the previous experiment that provided only binary feedback. 

\subsubsection{Fulfilled Singleton Predicates and Pass Rate (Figure \ref{fig:tool_self_imp})} 
The missed predicates follow a similar pattern as in the previous experiment (Figure \ref{fig:self_imp}), except that all models increased the number of generated solutions with fulfilled predicates. 

\subsubsection{Rate of Passed Test Cases (Table \ref{tab:test-rate-setups})}
\begin{figure*}[t]
\centering
  \includegraphics[width=18cm]{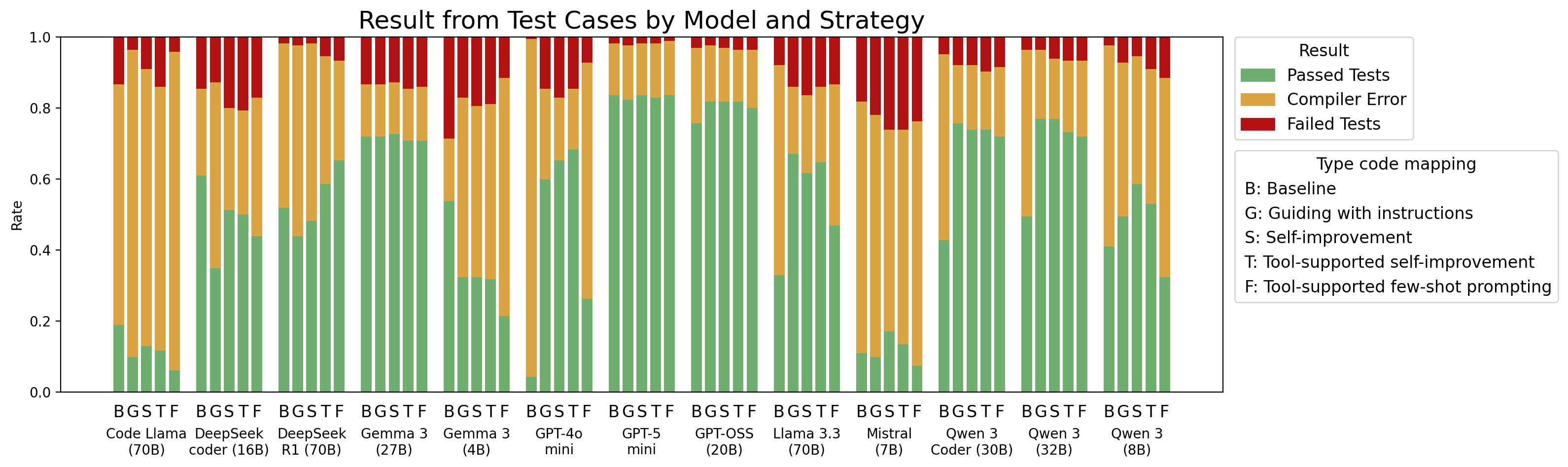} 
  \caption{The rate of passed test for all models in all experiments. Green represents passed tests, yellow represents generated solutions where the tests did not run because of compiler errors and red represents failed tests.}
  \label{fig:pass_rate}
\end{figure*}

The rate of passed test cases changed for most models compared to the baseline. However, each model was affected differently. For some models, Llama3.3, Qwen3 Coder, GPT-4o Mini, and Qwen3 (32B), there was a significant improvement. But for models such as Gemma3 (4B) and DeepSeek Coder there was instead a significant negative effect. Compared to the previous experiment, Guiding with instructions, the test pass rate was similar across all models except DeepSeek R1, which achieved a higher test pass rate with Tool-supported self-improvement, see Figure \ref{fig:pass_rate}.    

\subsection{Tool-supported few-shot prompting}
\subsubsection{Singleton Score (Table \ref{tab:singleton-scores})}
Six models were able to implement Singleton in all of the generated code.
Three more models achieved an average Singleton score of 95.3 or more on the generated code. DeepSeek r1 and Code Llama achieved their best scores with this strategy, achieving an average Singleton score of 85.4 and 43.3, respectively. However, Gemma3 (4B) did not benefit from this strategy and did only achieved an average Singleton score of 40.7. 

\subsubsection{Fulfilled Singleton Predicates and Pass Rate (Figure \ref{fig:tool_few_shot})}
For Gemma3 (4B) there was an increase in missing instances of all three predicates. However, Private Constructor was the one most affected, and was present in 20.7\% of the generated code. 


\subsubsection{Rate of Passed Test Cases (Table \ref{tab:test-rate-setups})}
Only five models achieved a significantly better test pass rate with this strategy, and four achieved a significantly lower pass rate than the baseline. This was not the most optimal strategy for any model except DeepSeek r1. DeepSeek r1 achieved the highest test pass rate and the highest average Singleton Score in the generated code among the experiment trials.
The drop in passed test rate, compared to the three other experiment trials for Code Llama, GPT-4o Mini, Llama 3.3 and Qwen3 can be seen in Figure \ref{fig:pass_rate}.

The addition of the Singleton examples did not improve the LLM's ability to implement the Singleton design pattern into the generated code compared to solely relying on feedback about which predicates were not fulfilled. However, for some LLMs the test pass rate was significantly lower than for the previous experiment with Tool-supported self-improvement. 

\begin{tcolorbox}[colback=beige, colframe=beige, boxrule=0pt,
                  left=8pt,right=8pt,top=8pt,bottom=8pt, breakable]
\textbf{Answer to RQ2:} Automated feedback strongly and consistently increases the LLM's ability to implement Singleton in the generated code. It can push strong models toward near-perfect or perfect predicate fulfillment and recuse common missing predicates. Predicate-specific feedback provides an additional performance boost compared to binary feedback. However, the degree of improvement remains model-specific, with weaker models benefiting less.

\end{tcolorbox}
\begin{tcolorbox}[colback=beige, colframe=beige, boxrule=0pt,
                  left=8pt,right=8pt,top=8pt,bottom=8pt, breakable]

\textbf{Answer to RQ3:} LLMs can be guided to generate Singleton classes in Java to a high degree. Often, it is nearly perfect with detailed feedback on missing predicates. In many cases, it even boosts the functional correctness of the generated code. However, the effect on functionality is not guaranteed and model-specific. 

\end{tcolorbox}

\section{Threats to Validity}

\textbf{Internal Validity}.
We used the predefined test cases in HumanEval-X to determine whether the generated code implemented the correct functionality. However, using test cases to evaluate functionality has limitations and is not an optimal approach. The test cases must be written to reflect the requirements of the problem definition. However, it is impossible to cover all scenarios with tests, and therefore, there is always a risk that the test cases used are inadequate for the given task. By using a well-established dataset, we mitigate the risk of inadequate testing.

\textbf{External Validity}.
The dataset used in each experiment is small, containing only 164 tasks. However, it included test cases to evaluate the code's functionality, which helped determine how well the LLMs would implement Singleton in the generated code without compromising functionality. All experiments included only tasks for which the desired solutions should be written in Java, so the findings cannot be generalized to other languages. 

Each model was only allowed a single attempt for each task with each strategy. This made the evaluation easy and straightforward; however, due to the nature of LLM models and their inherent randomness, we could likely observe slightly different results if we were to re-run all experiments. To estimate the difference, we re-ran all models for the iteration of experiment 1. Even though a small difference could be seen, it was not considered significant enough to rerun all experiments. 

\textbf{Construct Validity}.
In the experiments, the same prompt was used for all models, but LLMs are sensitive to prompt content and structure; therefore, the prompt used in the experiment is likely not optimal for all LLMs. However, we used the same prompt for all models, which was not optimized for any of them but rather chosen for its simplicity. An alternative approach would be to optimize the prompt for each LLM; however, in this study, we focused on including multiple LLMs rather than optimizing the strategies for each LLM. 

In the experiments, we used a script to check whether the generated solution satisfies the requirements for a Singleton class. This script was written to evaluate whether each predicate for a Singleton was present in the generated solution. As defined in this paper, a Singleton can be identified by this script based on fulfilled predicates. There are, however, multiple definitions of what a Singleton is and how it can be structured, and in this study, we evaluate only a single definition. 

The dataset used included 164 tasks, but none of the solutions to these tasks actually require implementing Singleton to be correct. In other words, through our experiments, we are encouraging the models to implement redundant design patterns. We are not evaluating whether the design patterns were implemented appropriately. However, the scope of this study was to experiment with which strategy would help guide the models towards implementing Singleton; therefore, the absence of tasks requiring Singleton was seen as a positive rather than a negative aspect.

\section{Future Work}

Future work will focus on including more types and a wider range of design pattern variations. For our next study on design patterns, we will explore how to guide LLMs to generate more abstract behavioral patterns, such as Strategy and Observer. Many design patterns, such as Singleton, also come in variations that differ somewhat in implementation. We will look into how to guide LLMs to generate the most suitable version for integration with a broader codebase.

Many previous studies have focused on recognizing design patterns in code and, as we have done here, generating code following design patterns. The next step in this process would be to evaluate how we can automatically recognize when it is appropriate to generate code that follows a specific design pattern, based on features in a given codebase. It will also be of interest to evaluate methods for generating code that follows or makes use of already defined design patterns in a codebase. Even if LLMs can detect and generate according to design patterns, their ability to use them correctly is unknown.




\section{Conclusion}

Across four experiments, guiding LLMs to consistently implement the Singleton design pattern in generated solutions increases pattern adherence, but the impact on functional correctness varies substantially across models. 
Although all evaluated models are capable of generating Singleton implementations through direct instruction, the correctness of these implementations is strongly model-dependent. 
Automated feedback loops further and more consistently enhance Singleton compliance, frequently correcting common missing predicates such as the Global Access Point and pushing stronger models toward correct functional implementations. Overall, no single guiding strategy is optimal: the effectiveness of direct instruction, automated feedback and predicate-level guide depends on the models' capabilities.




\section*{Data Availability}
All data used for evaluation, the results from these evaluations, together with the code used for running the trials and evaluation, is accessible here:  https://doi.org/10.5281/zenodo.18262240

\section*{Acknowledgment}
The Swedish Research Council partially funded this study under grant number 2024-04687. 

\bibliographystyle{ACM-Reference-Format}
\bibliography{ref}

@String{Computer = "{IEEE} Computer" }

@String{Psychometrika = "Psychometrika" }

@String{Springer = "Springer-Verlag" }

@InProceedings{singleton_def,
author="Stencel, Krzysztof
and W{\k{e}}grzynowicz, Patrycja",
editor="Meersman, Robert
and Tari, Zahir
and Herrero, Pilar",
title="Implementation Variants of the Singleton Design Pattern",
booktitle="On the Move to Meaningful Internet Systems: OTM 2008 Workshops",
year="2008",
publisher="Springer Berlin Heidelberg",
address="Berlin, Heidelberg",
pages="396--406",
isbn="978-3-540-88875-8"
}

@book{gamma1995design,
  title={Design patterns: elements of reusable object-oriented software},
  author={Gamma, Erich},
  volume={431},
  year={1995},
  publisher={Addison-Wesley}
}

@book{buschmann2007pattern,
  title={Pattern-oriented software architecture, on patterns and pattern languages},
  author={Buschmann, Frank and Henney, Kevlin and Schmidt, Douglas C},
  year={2007},
  publisher={John wiley \& sons}
}

@INPROCEEDINGS{GPTsurvey,
  author={Jahić, Jasmin and Sami, Ashkan},
  booktitle={2024 IEEE 21st International Conference on Software Architecture Companion (ICSA-C)}, 
  title={State of Practice: LLMs in Software Engineering and Software Architecture}, 
  year={2024},
  volume={},
  number={},
  pages={311-318},
  doi={10.1109/ICSA-C63560.2024.00059}}

@article{KIM2025112519,
title = {Comparative analysis of design pattern implementation validity in LLM-based code refactoring},
journal = {Journal of Systems and Software},
volume = {230},
pages = {112519},
year = {2025},
issn = {0164-1212},
doi = {https://doi.org/10.1016/j.jss.2025.112519},
url = {https://www.sciencedirect.com/science/article/pii/S0164121225001876},
author = {Dae-Kyoo Kim}
}

@INPROCEEDINGS{DoLLMunderstandDP,
  author={Pan, Zhenyu and Song, Xuefeng and Wang, Yunkun and Cao, Rongyu and Li, Binhua and Li, Yongbin and Liu, Han},
  booktitle={2025 IEEE/ACM International Workshop on Large Language Models for Code (LLM4Code)}, 
  title={Do Code LLMs Understand Design Patterns?}, 
  year={2025},
  volume={},
  number={},
  pages={209-212},
  doi={10.1109/LLM4Code66737.2025.00031}}

@INPROCEEDINGS{TransDPR,
  author={Pandey, Sushant Kumar and Staron, Miroslaw and Horkoff, Jennifer and Ochodek, Mirosław and Mucci, Nicholas and Durisic, Darko},
  booktitle={2023 ACM/IEEE International Symposium on Empirical Software Engineering and Measurement (ESEM)}, 
  title={TransDPR: Design Pattern Recognition Using Programming Language Models}, 
  year={2023},
  volume={},
  number={},
  pages={1-7},
  doi={10.1109/ESEM56168.2023.10304862}}

@article{selfRefined,
  title={Self-refine: Iterative refinement with self-feedback},
  author={Madaan, Aman and Tandon, Niket and Gupta, Prakhar and Hallinan, Skyler and Gao, Luyu and Wiegreffe, Sarah and Alon, Uri and Dziri, Nouha and Prabhumoye, Shrimai and Yang, Yiming and others},
  journal={Advances in Neural Information Processing Systems},
  volume={36},
  pages={46534--46594},
  year={2023}
}

@inproceedings{chen2024iterative,
  title={Iterative translation refinement with large language models},
  author={Chen, Pinzhen and Guo, Zhicheng and Haddow, Barry and Heafield, Kenneth},
  booktitle={Proceedings of the 25th Annual Conference of the European Association for Machine Translation (Volume 1)},
  pages={181--190},
  year={2024}
}

@inproceedings{NEURIPS2023_1b44b878,
 author = {Shinn, Noah and Cassano, Federico and Gopinath, Ashwin and Narasimhan, Karthik and Yao, Shunyu},
 booktitle = {Advances in Neural Information Processing Systems},
 editor = {A. Oh and T. Naumann and A. Globerson and K. Saenko and M. Hardt and S. Levine},
 pages = {8634--8652},
 publisher = {Curran Associates, Inc.},
 title = {Reflexion: language agents with verbal reinforcement learning},
 volume = {36},
 year = {2023}
}

@inproceedings{NEURIPS2023_7cc1005e,
 author = {Kim, Geunwoo and Baldi, Pierre and McAleer, Stephen},
 booktitle = {Advances in Neural Information Processing Systems},
 editor = {A. Oh and T. Naumann and A. Globerson and K. Saenko and M. Hardt and S. Levine},
 pages = {39648--39677},
 publisher = {Curran Associates, Inc.},
 title = {Language Models can Solve Computer Tasks},
 volume = {36},
 year = {2023}
}

@article{chen2023teaching,
  title={Teaching large language models to self-debug},
  author={Chen, Xinyun and Lin, Maxwell and Sch{\"a}rli, Nathanael and Zhou, Denny},
  journal={arXiv preprint arXiv:2304.05128},
  year={2023}
}

@inproceedings{peng2025perfcodegen,
  title={Perfcodegen: Improving performance of llm generated code with execution feedback},
  author={Peng, Yun and Gotmare, Akhilesh Deepak and Lyu, Michael R and Xiong, Caiming and Savarese, Silvio and Sahoo, Doyen},
  booktitle={2025 IEEE/ACM Second International Conference on AI Foundation Models and Software Engineering (Forge)},
  pages={1--13},
  year={2025},
  organization={IEEE}
}

@inproceedings{NEURIPS2020_1457c0d6,
 author = {Brown, Tom and Mann, Benjamin and Ryder, Nick and Subbiah, Melanie and Kaplan, Jared D and Dhariwal, Prafulla and Neelakantan, Arvind and Shyam, Pranav and Sastry, Girish and Askell, Amanda and Agarwal, Sandhini and Herbert-Voss, Ariel and Krueger, Gretchen and Henighan, Tom and Child, Rewon and Ramesh, Aditya and Ziegler, Daniel and Wu, Jeffrey and Winter, Clemens and Hesse, Chris and Chen, Mark and Sigler, Eric and Litwin, Mateusz and Gray, Scott and Chess, Benjamin and Clark, Jack and Berner, Christopher and McCandlish, Sam and Radford, Alec and Sutskever, Ilya and Amodei, Dario},
 booktitle = {Advances in Neural Information Processing Systems},
 editor = {H. Larochelle and M. Ranzato and R. Hadsell and M.F. Balcan and H. Lin},
 pages = {1877--1901},
 publisher = {Curran Associates, Inc.},
 title = {Language Models are Few-Shot Learners},
 volume = {33},
 year = {2020}
}

@INPROCEEDINGS{10298329,
  author={Gao, Shuzheng and Wen, Xin-Cheng and Gao, Cuiyun and Wang, Wenxuan and Zhang, Hongyu and Lyu, Michael R.},
  booktitle={2023 38th IEEE/ACM International Conference on Automated Software Engineering (ASE)}, 
  title={What Makes Good In-Context Demonstrations for Code Intelligence Tasks with LLMs?}, 
  year={2023},
  volume={},
  number={},
  pages={761-773},
  doi={10.1109/ASE56229.2023.00109}}

@article{mcnemar1947note,
  title={Note on the sampling error of the difference between correlated proportions or percentages},
  author={McNemar, Quinn},
  journal={Psychometrika},
  volume={12},
  number={2},
  pages={153--157},
  year={1947},
  publisher={Springer-Verlag}
}

@article{SERGEYUK2025107610,
title = {Using AI-based coding assistants in practice: State of affairs, perceptions, and ways forward},
journal = {Information and Software Technology},
volume = {178},
pages = {107610},
year = {2025},
issn = {0950-5849},
doi = {https://doi.org/10.1016/j.infsof.2024.107610},
url = {https://www.sciencedirect.com/science/article/pii/S0950584924002155},
author = {Agnia Sergeyuk and Yaroslav Golubev and Timofey Bryksin and Iftekhar Ahmed}
}

@inproceedings{tian2025fixing,
  title={Fixing Large Language Models' Specification Misunderstanding for Better Code Generation},
  author={Tian, Zhao and Chen, Junjie and Zhang, Xiangyu},
  booktitle={2025 IEEE/ACM 47th International Conference on Software Engineering (ICSE)},
  pages={645--645},
  year={2025},
  organization={IEEE Computer Society}
}

@article{Samo_Ali_Memon_Abbasi_Koondhar_Dahri_2024, title={Fine-Tuning Mistral 7b Large Language Model For Python Query Response And Code Generation: A Parameter Efficient Approach}, volume={12}, url={https://vfast.org/journals/index.php/VTCS/article/view/1885}, DOI={10.21015/vtcs.v12i1.1885}, abstractNote={&amp;lt;p&amp;gt;&amp;lt;em&amp;gt;The research delves into the concept of fine-tuning and its unique application to Python queries and code generation. This process involves adjusting the model&amp;#039;s parameters to make it more proficient in responding to Python-only queries and generating corresponding code. It underscores the untapped potential of fine-tuning large language models and the significance of combining Parameter-Efficient fine-tuning with quantization to reduce memory usage. This was achieved by the pre-trained language model’s fine-tuning and meticulously evaluating and contrasting it with its base model. Notably, the model was fine-tuned to proficiently respond to Python-only queries and generate corresponding code, a novel and intriguing application of fine-tuning. &amp;lt;/em&amp;gt;&amp;lt;em&amp;gt;We utilized Mistral7B-Instruct version 0.2 as our base large language model for fine-tuning. The dataset, sourced from Kaggle, was a collection of Python Question-Answer pairs. Before fine-tuning, we meticulously cleaned and organized the dataset, ensuring its quality by arranging it in descending order based on their rankings. This rigorous and thorough approach instills confidence in the reliability of our results. &amp;lt;/em&amp;gt;&amp;lt;em&amp;gt;Our research shows that the fine-tuned model outperformed the base Mistral7B Instruct version 0.2 model in BERTScore and demonstrated a significant performance boost when compared with the HumanEval metric. This clear and substantial improvement further affirms the effectiveness of our approach. &amp;lt;/em&amp;gt;&amp;lt;em&amp;gt;Our research is a step forward in the realm of large language models, specifically in the coding sphere. It showcases a substantial improvement in understanding Python queries and generating code snippets. This has profound implications for the current trajectory of Natural Language Generation and Generative AI fields. Our findings act as a pivotal catalyst for further progress.&amp;lt;/em&amp;gt;&amp;lt;/p&amp;gt;}, number={1}, journal={VAWKUM Transactions on Computer Sciences}, author={Samo, Hassan and Ali, Kashif and Memon, Muniba and Abbasi, Faheem Ahmed and Koondhar, Muhammad Yaqoob and Dahri, Kamran}, year={2024}, month={Jun.}, pages={205–217} }

@misc{yang2025qwen3technicalreport,
      title={Qwen3 Technical Report}, 
      author={An Yang and Anfeng Li and Baosong Yang and Beichen Zhang and Binyuan Hui and Bo Zheng and Bowen Yu and Chang Gao and Chengen Huang and Chenxu Lv and Chujie Zheng and Dayiheng Liu and Fan Zhou and Fei Huang and Feng Hu and Hao Ge and Haoran Wei and Huan Lin and Jialong Tang and Jian Yang and Jianhong Tu and Jianwei Zhang and Jianxin Yang and Jiaxi Yang and Jing Zhou and Jingren Zhou and Junyang Lin and Kai Dang and Keqin Bao and Kexin Yang and Le Yu and Lianghao Deng and Mei Li and Mingfeng Xue and Mingze Li and Pei Zhang and Peng Wang and Qin Zhu and Rui Men and Ruize Gao and Shixuan Liu and Shuang Luo and Tianhao Li and Tianyi Tang and Wenbiao Yin and Xingzhang Ren and Xinyu Wang and Xinyu Zhang and Xuancheng Ren and Yang Fan and Yang Su and Yichang Zhang and Yinger Zhang and Yu Wan and Yuqiong Liu and Zekun Wang and Zeyu Cui and Zhenru Zhang and Zhipeng Zhou and Zihan Qiu},
      year={2025},
      eprint={2505.09388},
      archivePrefix={arXiv},
      primaryClass={cs.CL},
      url={https://arxiv.org/abs/2505.09388}, 
}

@ARTICLE{10952968,
  author={Staron, Miroslaw and Abrahão, Silvia},
  journal={IEEE Software}, 
  title={Exploring Generative AI in Automated Software Engineering}, 
  year={2025},
  volume={42},
  number={3},
  pages={142-145},
  doi={10.1109/MS.2025.3533754}}

@misc{wang2026codeflowbenchmultiturniterativebenchmark,
      title={CodeFlowBench: A Multi-turn, Iterative Benchmark for Complex Code Generation}, 
      author={Sizhe Wang and Zhengren Wang and Dongsheng Ma and Yongan Yu and Rui Ling and Zhiyu Li and Feiyu Xiong and Wentao Zhang},
      year={2026},
      eprint={2504.21751},
      archivePrefix={arXiv},
      primaryClass={cs.SE},
      url={https://arxiv.org/abs/2504.21751}, 
}

@misc{bi2025gptossgoodcomprehensiveevaluation,
      title={Is GPT-OSS Good? A Comprehensive Evaluation of OpenAI's Latest Open Source Models}, 
      author={Ziqian Bi and Keyu Chen and Chiung-Yi Tseng and Danyang Zhang and Tianyang Wang and Hongying Luo and Lu Chen and Junming Huang and Jibin Guan and Junfeng Hao and Xinyuan Song and Junhao Song},
      year={2025},
      eprint={2508.12461},
      archivePrefix={arXiv},
      primaryClass={cs.CL},
      url={https://arxiv.org/abs/2508.12461}, 
}

@inproceedings{nguyen2015sospa,
  title={Sospa: A system of security design patterns for systematically engineering secure systems},
  author={Nguyen, Phu H and Yskout, Koen and Heyman, Thomas and Klein, Jacques and Scandariato, Riccardo and Le Traon, Yves},
  booktitle={2015 ACM/IEEE 18th International Conference on Model Driven Engineering Languages and Systems (MODELS)},
  pages={246--255},
  year={2015},
  organization={IEEE}
}


\end{document}